# Electronic transport properties of ternary Cd$_{1-x}$Zn$_x$S nanowire network

**Daeha Joung[1,2], M. Arif[1], S. Biswas[1], S. Kar[1], S. Santra[1,3,4] and Saiful I. Khondaker [1,2*]**
[1]Nanoscience Technology Center, [2]Department of Physics, [3]Department of Chemistry and [4]Biomolecular Science Center, University of Central Florida, 12424 Research Parkway, Suite 400, Orlando, FL 32826

* To whom correspondence should be addressed. E-mail: saiful@mail.ucf.edu

**Abstract**

We present electronic transport characteristics of ternary alloy Cd$_{1-x}$Zn$_x$S nanowire networks in the dark and under white light illumination. Compared to the negligible dark current, we observed a photocurrent enhancement up to 4 orders of magnitude at intensity of 460 mW/cm$^2$. The time constant of the dynamic photoresponse is ~5 sec. The current – voltage characteristics at different intensities show Ohmic behavior at low bias and space charge limited conduction (SCLC) at higher bias voltages. The SCLC behavior and slow time response indicate that the charge transport is dominated by tunneling at the percolating inter-nanowire junctions.

(Some figures in this article are in color only in the electronic version)

**1. Introduction:**

Chemically synthesized semiconductor nanocrystals have attracted tremendous interest because of the ability to tune their band gap through variation of their sizes and shapes. These nanocrystalline materials, specially that work in the visible range, have great potential for applications in optoelectronic devices such as solar cells, light emitting diodes, lasers, waveguides and photodetectors [1-5]. In addition, their ease of processibility from solution is of great advantage as they can cover large areas at low cost and are compatible with flexible substrates. Many studies have been done on semiconductor quantum dots [4-8] and quantum dot polymer composites [9-11]. However, it is believed that materials that are one dimensional (nanorods, nanowires) can provide added benefit as they make a percolating pathway for charge transport rather easily compared to zero dimensional quantum dots. There have been several reports on the optoelectronic studies of binary nanowires (NWs) [1-3, 12-15]. For example, Huynh et al used CdSe nanorods for making organic-inorganic hybrid solar cells [2]. Kind et al used ZnO NWs for ultraviolet photodetectors and optical switches [3]. Nonetheless, tuning the bandgap by varying the diameter and aspect ratio of one-dimensional NWs is extremely challenging. An alternative route of tuning the bandgaps of NWs is to engineer the relative composition of the materials to fabricate ternary alloy NWs. Despite the obvious advantage of ternary nanowires over binary NWs, there are only a few reports on the growth and optical properties of several ternary NW systems [16-19], and no reports on the optoelectronic investigation of any ternary NWs. Study of charge transport and photoresponse properties of ternary NW network film is important in order to determine whether this material could be useful for optoelectronic applications.

In this paper, we report on the first optoelectronic transport characteristics of ternary alloy Cd$_{1-x}$Zn$_x$S NW network devices. The NW films were prepared by drop-casting the NWs from solution on inter-digitated gold source and drain electrodes fabricated on Si/SiO$_2$ substrate.

The current – voltage ($I – V$) characteristics were measured in the dark and under illumination from white light of different intensity using a xenon lamp (oriel solar simulator). The dark current was negligible, while the current under illumination increases almost linearly upon increasing the intensity of the light. We observed an enhancement of photocurrent up to four orders of magnitude compared to the dark current at an intensity of 460 mW/cm$^2$ and a bias voltage of 15 V. The time constant of dynamic photoresponse is slow (~ 5 sec). The *I-V* curve at different intensities show Ohmic behavior at low bias and space charge limited conduction (SCLC) at high bias voltage. The space charge limited conduction along with slow time response show that the charge conduction is dominated by the carrier tunneling through inter-NW junctions. This study will have significant impact on the use of $Cd_{1-x}Zn_xS$ NWs in optoelectronics.

## 2. Experimental details:

The $Cd_{1-x}Zn_xS$ NWs were grown following our recently developed technique [20]. Briefly an appropriate amount of zinc acetate [$(CH_3COO)_2Zn$, $2H_2O$], cadmium acetate [$(CH_3COO)_2Cd$, $4H_2O$], and thiourea (**tu**, $NH_2CSNH_2$) were placed in the Teflon-lined chamber, which was then filled with an ethylenediamine (**en**, $NH_2CH_2CH_2NH_2$) and water mixture with 2:1 volume ratio up to 80% of its volume. The closed chamber was placed inside a pre-heated oven at 175 °C for 8 hours and then cooled at room temperature. The resulting precipitates were filtered off and rinsed several times in water and ethanol, then dried in vacuum at room temperature for 6 hours to obtain the final powder product. Figure 1(a) shows the representative transmission electron microscopy image of $Cd_{1-x}Zn_xS$ NWs with x=0.75. The NW is single crystalline and the average diameter of the NWs is between 8 to 15 nm which remains almost the same with the variation in the composition of alloy NWs, while the length of the NWs can be up to 500 nm [20]. In this paper we used three different NW samples with x = 0, 0.25, 0.75. The band gap energy of the NWs increases from 2.37 eV to 3.13 eV for the increase in the x value from 0 to 0.75 as revealed from the UV-vis absorption spectra in figure 1(b). As the exact amount of $Cd^{2+}$ and $Zn^{2+}$ ions in the $Cd_{1-x}Zn_xS$ lattice varied from the added amount of precursors used in the synthesis [20], band gap energy of the materials differed from the theoretical value (using analytical expression $E_g = 2.337 + 0.72x + 0.563x^2$ for x=0.25, 0.75 [21].

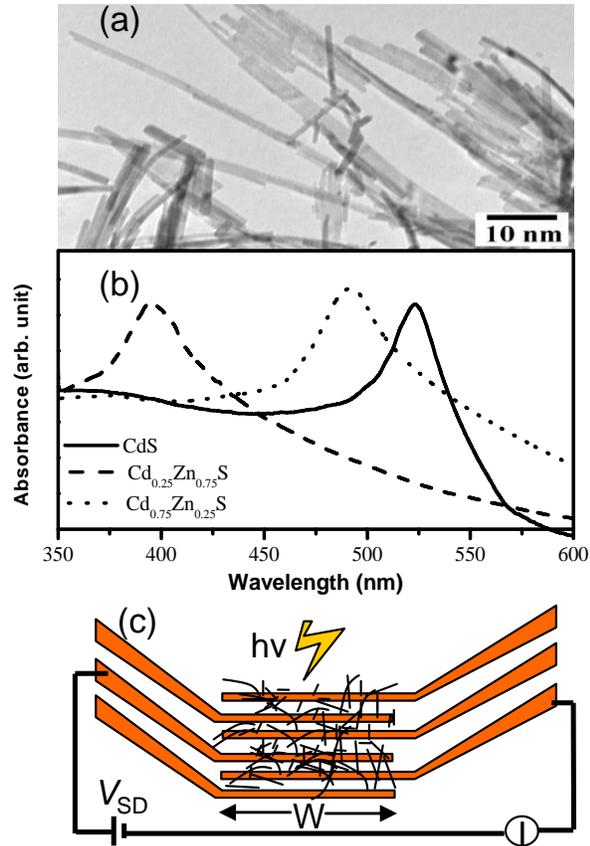

**Figure 1.** (a) Transmission electron micrograph image of Cd $_{0.25}$Zn $_{0.75}$S NWs. (b) UV -vis absorption spectra of NWs with different composition of Cd and Zn. (c) Cartoon of the device with transport measurement setup.

The inter-digitated array electrodes were fabricated using standard optical lithography on a $SiO_2$ (250 nm thick)/Si wafer followed by deposition of Cr (5 nm)/Au (25 nm) and lift off in acetone. The channel width was W=200 μm while the effective channel lengths L were 5, 9 and 17 μm. The width of each electrode was 5 μm. The NW film was made using a standard drop casting method. The NWs in chloroform were ultra-sonicated for 10 minutes prior to drop casting. A small drop (1 μl) of solution was drop casted in the center of the sample containing source and drain electrodes. The sample was left to dry at atmosphere before electrical characterization at a probe station. The film thickness was about 175 nm. Figure 1(c) shows a schematic of our devices including transport measurement setup. The room temperature dc transport measurements of the NW films were carried out at ambient condition using a standard two-probe technique both in dark and under illumination by a xenon lamp from a solar-light simulator (Oriel 96000) with a maximum power of 300 W. The light source was placed 10 cm above the sample and the intensity was measured with a calibrated silicon photodiode. The intensity of the light source was varied from 0 to 464 mW/cm$^2$. The *I-V* characteristics in the dark and under illumination were measured using Keithley 2400 source meter and a current preamplifier (DL instruments: Model 1211) capable of measuring sub pA signal. These instruments were interfaced with LabView for data collection.

### 3. Results and Discussions:

Figure 2(a) shows the I-V characteristics for $Cd_{0.25}Zn_{0.75}S$ film with channel length L=5 μm under white light illumination with intensities of 51, 102, 165, 236, 322, 373, and 464 mW/cm$^2$. The dark *I-V* curve is shown in the inset of figure 2(a). In the dark, the NW is highly insulating with a current of only 35 pA at 20 V bias. As the NW film was illuminated, the photocurrent is enhanced. The *I-V* curve shows nonlinear behavior with increasing bias voltage. At a fixed bias voltage, the photocurrent increases with increasing intensity. At the highest intensity of 464 mW/cm$^2$ in our measurements, the photocurrent ($I_{photo}$) at 15 V was 203 nA while the dark current ($I_{dark}$) was only 20 pA resulting in an enhancement of photocurrent ($I_{photo}/I_{dark}$) of ~10$^4$. In order to check whether there was any effect of temperature due to this high intensity, we monitored the temperature of the film using a thermocouple. The temperature of the film increased from 25$^o$C to 35$^o$C when illuminated at the highest intensity of 464 mW/cm$^2$. We then measured the dark I-V characteristics by heating the sample at 35$^o$C. There was no change in the dark I-V curve compared to the room temperature dark I-V curve due to this heating effect. In other words, the observed photoresponse of up to 4 orders of magnitude is not affected by a small change of temperature under light illumination. Similar enhancement of photocurrent under white light illumination has been observed for other samples with x=0, and x=0.75. The UV-VIS absorption data in figure 1(b) shows that the absorption changes from 350 nm for $Cd_{0.25}Zn_{0.75}S$ up to 600 nm for CdS. It can also be seen that the absorbance is almost the same for all the materials at peak wavelength. In our study, we used a white light from a solar simulator which covers all the wavelengths that the NW absorbs. Therefore similar photoresponse behavior under white light illumination is expected in our study.

Figure 2(b) shows the photocurrent versus intensity for the same sample presented in figure 2(a), with different electrode separations. As the intensity is increased, the photocurrent is increased from 0.01 to 10.96 nA for 17 μm, 0.02 to 60.15 nA for 9 μm, and 0.02 to 203 nA for 5 μm electrode separation. The dependence of photocurrent on intensity can be described using a power law relation; $I = A \cdot P^\theta$ where $A$ is a constant, $P$ is the power of illumination and $\theta$ is the exponent characterizing charge transport in the semiconductor film. The circles are experimental points and the solid lines are fits to the power law equation. From the fit, we obtained $A = 1.4$, $\theta = 0.82$ for 5 μm, $A = 0.43$, $\theta = 0.60$ for 9 μm, and $A = 0.023$, $\theta = 0.86$ for 17 μm channel length. The value of $\theta < 1$ indicates the presence of disorder in our film, probably due to the charge carrier tunneling at the inter-NW junctions. Additionally, the process of electron-hole generation, trapping and recombination within the network can also cause $\theta < 1$ [22]. A small variation of $\theta$ for different electrode spacing could be due to the non uniformity of the film over long distance as the film was made by drop casting. However, the value of $\theta$ between 0.5 and 1 is consistent with that reported in other nanostructures such as ZnO NWs and CdS nanobelts [3, 12].

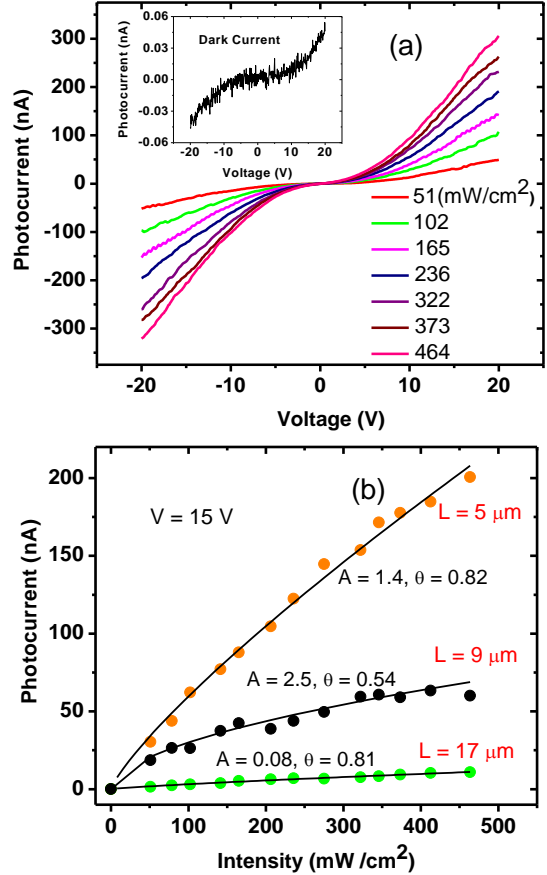

**Figure 2.** (a) Typical $I-V$ characteristics of a $Cd_{0.25}Zn_{0.75}S$ NW film with electrode separation $L$=5 μm, under white-light illumination of different intensity. Inset: $I-V$ curve in the dark. (b) Photocurrent versus light intensity of $Cd_{0.25}Zn_{0.75}S$ film for L = 5, 9 and 17 μm. The circles are the experimental points and the solid lines are fits using $I = AP^\theta$

Figure 3(a) shows the photocurrent as a function of time ($t$) for the same sample presented in figure 2(a) for several cycles when the light was turned on and off with 20 second intervals starting at $t$ =10 seconds. The bias voltage was 10 V and the light intensity was 373 mW/cm$^2$. It can be seen from this figure that time response for photocurrent is very slow. We found that our data can be well described using the equation $I(t) = A_1 \exp(-t/\tau_1) + A_2 \exp(-t/\tau_2) + I_0$, where $A_1$ and $A_2$ are scaling constants, $\tau_1$ and $\tau_2$ are two characteristic time constants [23]. Figure 3(b) shows the raising part of the photocurrent after the light was turned on while figure 3(c) shows decaying part after the light was turned off along with the fit to the above equation. The open circles are the experimental data points while the solid lines are the fits. It can be seen that a very good fit is obtained with $\tau_1$ = 0.3 sec, $\tau_2$ = 5.0 sec for the raising part and $\tau_1$ = 0.2 sec, $\tau_2$ = 4.3 sec for the decaying part. We would like to note that, we also attempted to fit our data with a single exponential,

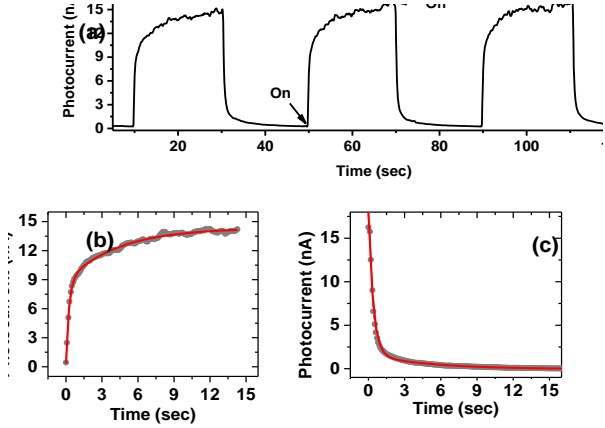
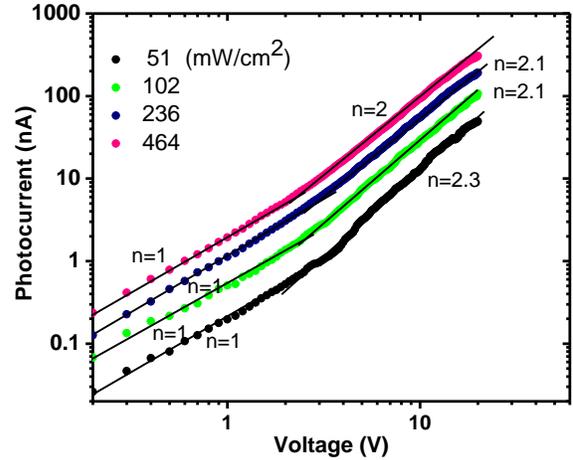

**Figure 3.** Time response of the photocurrent for $Cd_{0.25}Zn_{0.75}S$ film. (a) Photocurrent versus time for several cycles as the light was turned on and off. (b) Photocurrent along with exponential fits when the light was turned on and (c) when light was turned off.

**Figure 4.** Photocurrent versus voltage plotted in log-log scale for $Cd_{0.25}Zn_{0.75}S$ film. The squares are data points and the lines are fits using $I \propto V^n$. At low bias (<2V) the conduction is Ohmic and at high bias (>2V) the conduction is space charge limited (SCLC).

$I(t) = I_0 + A\exp(-t/\tau)$ with only one time constant $\tau$, however the data did not fit well. The shorter time $\tau_1$ could be related to charge carrier generation when the light was switched on while the longer time $\tau_2$ could be related to the diffusion of charge carriers through the interpenetrating NW network where carrier needs to tunnel from one NW to the next through the NW junction. Charge carrier trapping and recombination in the network may also play a role in slow time response.

In order to further understand the charge conduction mechanism in our NW network, we have plotted the $I-V$ curves with intensities 51, 102, 236, and 464 mW/cm$^2$ in log-log scale which is shown in figure 4. The dotted symbols are the data points and the solid lines are a fit using $I \propto V^n$. It can be seen from this figure that at low biases (<2 V) $n = 1$, which implies the the conduction is Ohmic. However at higher bias voltages ( >2 V) $n$ varies from 2 to 2.3 (error bar ±0.2) with different light intensities. A value of $n \sim 2$ is characteristic of a SCLC behavior. Similar SCLC behavior has been observed in all our films. The presence of SCLC behavior in our CdZnS NW network provides further evidence that the charge transport is limited by tunneling through many interpenetrating NW junctions. Additionally disorder originating from surface defects, charge traps may also play a role which has been reported for GaAs, CdS and antimony sulfide NWs [13-16].

### 4. Conclusions:

In conclusion, we have investigated the electronic transport characteristics of ternary alloy $Cd_{1-x}Zn_xS$ NW network films in dark and under illumination. The enhancement of photocurrent was up to four orders of magnitude compared to the dark current. We observed that the photocurrent followed a power law behavior with intensity and that the time constant of dynamic photoresponse was large. The current-voltage characteristics showed nonlinear behavior at higher bias which can be explained using SCLC model. This indicates that the charge transport is dominated by the disorder created by inter-nanowire network.

This work has been partially supported by US NSF under grants ECCS 0748091 and ECCS 0801924 to SIK and EEC-056560 (NIRT) to SS.